\documentclass[
floatfix,
aps,
prl,
amsmath,
twocolumn,
superscriptaddress,
]{revtex4-1}

\usepackage{times}
\usepackage{amssymb}
\usepackage{latexsym}
\usepackage[dvips]{graphicx}
\usepackage{amsmath}
\usepackage{graphicx}
\usepackage{dcolumn}
\usepackage{amsfonts}
\usepackage{bm}
\usepackage{epsfig}
\newcommand{\be}{\begin{equation}}
\newcommand{\ee}{\end{equation}}
\newcommand{\bea}{\begin{eqnarray}}
\newcommand{\eea}{\end{eqnarray}}

\def\lb{\left [}
\def\rb{\right ]}

\def\edc{\epsilon_{j}}

\def\ttr{\tau}

\def\oc{\omega_{\mbox{\scriptsize {c}}}}

\def\rc{R_{\mbox{\scriptsize {c}}}}

\def\tpi{\tau_{\pi}}

\def\tq{\tau_{\mbox{\scriptsize {q}}}}
\def\tqa{\tau_{\mbox{\scriptsize {q1}}}}
\def\tqb{\tau_{\mbox{\scriptsize {q2}}}}
\def\mqa{\mu_{\mbox{\scriptsize {q1}}}}
\def\mqb{\mu_{\mbox{\scriptsize {q2}}}}

\def\ttr{\tau}

\def\m{m^\star}
\def\mqi{\mu_{\mbox{\scriptsize {q}},i}}

\newcommand{\req}[1]{Eq.\,(\ref{#1})}

\newcommand{\rfig}[1]{Fig.\,\ref{#1}}

\newcommand{\rref}[1]{Ref.\,\onlinecite{#1}}

\usepackage[english,american]{babel}
\begin{document}

\title{
Hall field-induced resistance oscillations in a tunable-density GaAs quantum well
}

\author{M.~A.~Zudov}
\affiliation{School of Physics and Astronomy, University of Minnesota, Minneapolis, Minnesota 55455, USA}
\author{I.~A.~Dmitriev}
\affiliation{Department of Physics, University of Regensburg, D-93040 Regensburg, Germany}
\affiliation{Max Planck Institute for Solid State Research, Heisenbergstrasse 1, D-70569 Stuttgart, Germany}
\affiliation{Ioffe Physical Technical Institute, 194021 St. Petersburg, Russia}
\author{B.~Friess}
\affiliation{Max Planck Institute for Solid State Research, Heisenbergstrasse 1, D-70569 Stuttgart, Germany}
\author{Q.~Shi}
\affiliation{School of Physics and Astronomy, University of Minnesota, Minneapolis, Minnesota 55455, USA}
\author{V.~Umansky}
\affiliation{Braun Centre for Semiconductor Research, Department of Condensed Matter Physics, \\Weizmann Institute of Science,
Rehovot 76100, Israel}
\author{K.~von~Klitzing}
\affiliation{Max Planck Institute for Solid State Research, Heisenbergstrasse 1, D-70569 Stuttgart, Germany}
\author{J.~Smet}
\affiliation{Max Planck Institute for Solid State Research, Heisenbergstrasse 1, D-70569 Stuttgart, Germany}

\begin{abstract}
We report on Hall field-induced resistance oscillations (HIRO) in a 60 nm-wide  GaAs/AlGaAs quantum well with an \emph{in situ} grown back gate, which allows tuning the carrier density $n$. 
At low $n$, when all electrons are confined to the lowest subband (SB1), the HIRO frequency, proportional to the product of the cyclotron diameter and the Hall field, scales with $n^{-1/2}$, as expected. 
Remarkably, population of the second subband (SB2) significantly enhances HIRO, while their frequency now scales as $n^{-1}$. 
We demonstrate that in this two-subband regime HIRO still originate solely from backscattering of SB1 electrons. 
The unusual density dependence occurs because the population of SB2 steadily increases, while that of SB1 remains essentially unchanged. 
The enhancement of HIRO manifests an unexpected, step-like increase of the quantum lifetime of SB1 electrons, which reaches a record value of 52 ps in the two-subband regime.
\end{abstract}

\received{7 August 2017}
\maketitle

Continuous developments \citep{pfeiffer:1989,umansky:1997,pfeiffer:2003,umansky:2009,umansky:2013,manfra:2014,watson:2015,gardner:2016} in the molecular beam epitaxy and heterostructure design of 2D electron systems (2DES) have led to discoveries of a plethora of novel phenomena, especially in the field of low-temperature magnetotransport. 
Apart from the extremely rich quantum Hall physics in strong magnetic fields \citep{klitzing:1980,tsui:1982b}, high-mobility 2DES display many prominent transport phenomena in low fields.
Two salient examples of such phenomena are microwave- (MIRO) \citep{zudov:2001b,ye:2001,mani:2002,zudov:2003,zudov:2014,karcher:2016} and Hall field-induced resistance oscillations (HIRO) \cite{yang:2002,bykov:2005c,zhang:2007a,bykov:2007,zhang:2008,hatke:2009c,hatke:2010a,hatke:2011a,shi:2014b,shi:2017b} which emerge when a 2DES is driven by microwave radiation and direct current, respectively. 

HIRO emerge due to elastic electron transitions between Landau levels, tilted by the Hall field, as a result of backscattering off short-range impurities \citep{yang:2002,vavilov:2007,lei:2007}.  
The probability of these transitions is maximized each time the Hall voltage drop across the cyclotron diameter matches an integer multiple of the cyclotron energy.
As a result, the differential resistivity acquires a $1/B$-periodic correction $\delta r$ which can be described by \citep{vavilov:2007}
\be
{\delta r}/{\rho_0} \approx ({16\tau}/{\pi\tau_{\pi}})\lambda^2 \cos (2\pi B_1/B) \,,
\label{eq.hiro}
\ee
where $\rho_0=\m/e^2 n\tau$ is the resistivity at zero magnetic field $B$, $\m \approx 0.07m_0$ is the effective mass, $n$ is the electron density, $\lambda = \exp(-\pi/\oc\tq)$ is the Dingle factor, $\oc = eB/\m$ is the cyclotron frequency, and $\tau, \tpi, \tq$ are transport, backscattering, and quantum lifetimes, respectively \citep{note:times}. 
The HIRO frequency (inverse period) $B_1$ is given by 
\be
\frac{B_1}B \equiv \frac{e E(2\rc)}{\hbar\oc}\Rightarrow B_1 = \sqrt{\frac{8\pi}{n}}\frac{\m}{e^2} j\,,
\label{eq.b1}
\ee
where $E = Bj/ne$ is the Hall field and $\rc = \hbar \sqrt{2\pi n}/eB$ is the cyclotron radius.

It is well known that in systems with several populated subbands, MIRO and HIRO often mix with magneto-intersubband oscillations (MISO) \cite{raichev:2008,wiedmann:2008,mamani:2009b,wiedmann:2010a,wiedmann:2010b,wiedmann:2011c,gusev:2011}.
However, it is also important to examine how MIRO and HIRO are affected by the population of the second subband in the absence of such mixing.
For example, capacitance measurements in a wide quantum well provided direct evidence of microwave-induced non-equilibrium redistribution of electrons between two subbands but no significant change in MIRO upon second subband population \cite{dorozhkin:2016b}. However, unlike MIRO, whose frequency $\m\omega/e$ ($\omega$ being the microwave frequency) is density-independent, the HIRO frequency in the single-subband regime $B_1$ scales as $n^{-1/2}$, which stems from the product of $\rc\propto \sqrt{n}$ and $E \propto n^{-1}$ [see \req{eq.b1}]. 
It is thus interesting to explore how HIRO evolve with the populations of the lowest (SB1) and the first excited (SB2) subbands and to detect HIRO contributions from each subband, which should provide access to their individual scattering rates.

In this Rapid Communication we report on Hall field-induced resistance oscillations in a density-tunable wide quantum well in which electrons form two parallel layers when SB2 becomes populated \citep{nuebler:2012,dorozhkin:2016,dorozhkin:2016b}. 
Remarkably, in the two-subband regime, we still observe only one set of HIRO but find that their frequency scales as $1/n$, seemingly in contradiction with \req{eq.b1}.
As demonstrated below, this finding indicates that HIRO still originate from backscattering of SB1 electrons only (whose density remains approximately constant in the two-subband regime), while the Hall field $E$ is produced by electrons in both subbands.
We further find that HIRO become markedly enhanced in the two-subband regime. 
This enhancement was not anticipated, but can be linked to an abrupt increase of the quantum scattering time of SB1 electrons. 
One possible reason for such an increase is an additional screening of the long-range scattering potential by SB2 electrons.
The absence of HIRO from SB2 electrons can be explained by their considerably shorter quantum scattering time. 
This conclusion is supported by a complementary analysis of MISO.

Our lithographically defined Hall bar (width $w = 0.4$ mm) sample was fabricated \citep{note:contacts} from a GaAs/AlGaAs heterostructure containing a 60 nm wide quantum well. 
The structure was modulation-doped on the top side using a short-period superlattice positioned 66 nm away from the quantum well. 
An additional \emph{in situ} grown doped quantum well, located 800 nm below the main quantum well, served as a gate to tune the carrier density \citep{note:structure,nuebler:2012}.
At zero gate voltage, the electron density and the mobility were $n \approx 1.1 \times 10^{11}$ cm$^{-2}$ and $\mu \approx 5.8 \times 10^6$ cm$^2$/Vs, respectively. 
The differential longitudinal resistivity $r=dV/dI$, where $V$ is the voltage drop between contacts positioned 1.26 mm apart along the Hall bar, and the Hall resistivity were measured while sweeping $B$ using a standard four-terminal lock-in technique (0.5 $\mu$A excitation) in a $^3$He cryostat at a base temperature of 0.3 K.

\begin{figure}[t]
\includegraphics{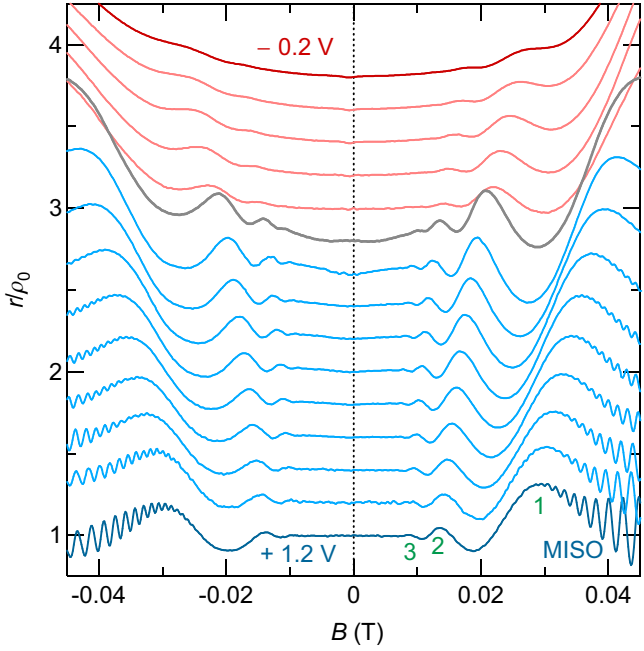}
\vspace{-0.1 in}
\caption{(Color online)
Differential resistivity $r$, normalized to $\rho_0$, vs $B$ at $I = 50$ $\mu$A and at gate voltages from $V_g = -0.2$ V (top) to $1.2$ V (bottom), in steps of 0.1 V.
The traces are vertically offset for clarity by 0.2 (bottom to top).
Numbers next to the bottom trace mark HIRO maxima near $B_1/B = 1, 2,$ and 3.
}
\vspace{-0.2 in}
\label{fig1}
\end{figure}

\begin{figure}[t]
\includegraphics{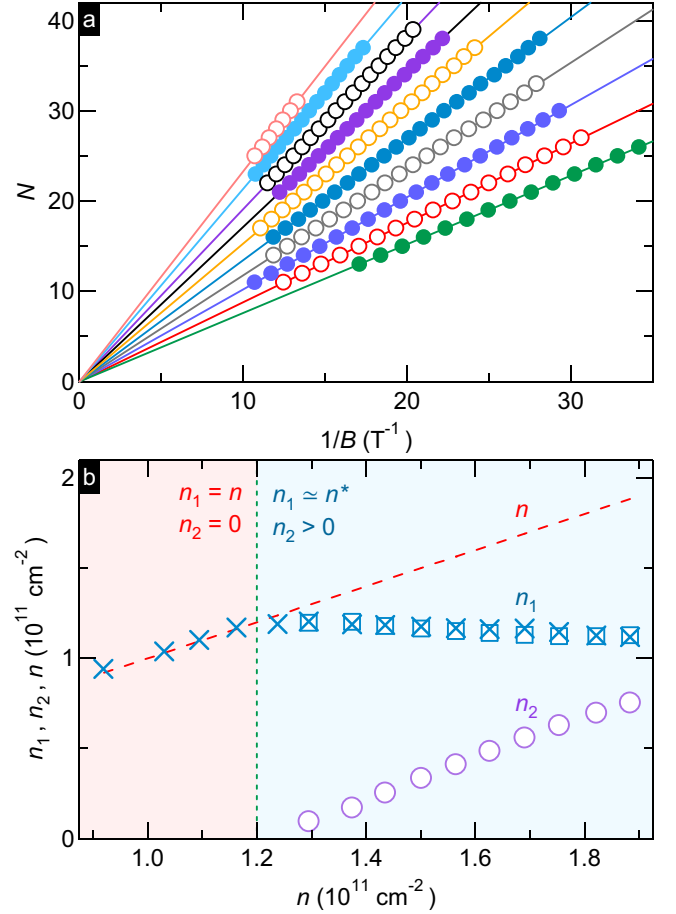}
\vspace{-0.1 in}
\caption{(Color online)
(a) MISO order $N$ vs $1/B$ for $V_g$ from $+1.2$ V (bottom) to $+0.3$ V (top), in steps of 0.1 V.
Lines are fits to \req{eq.N}.
(b) $n_1$ (squares) and $n_2$ (circles), obtained from \req{eq.n12}, and $n_1$ (crosses), obtained from \req{eq.b1} ($n < n^\star$) and \req{eq.b1m} ($n > n^\star$), as a function of $n$.
Dashed line represents $n = n_1 + n_2$.
}
\vspace{-0.2 in}
\label{fig2}
\end{figure}

In \rfig{fig1} we present the differential resistivity normalized to the zero-field resistivity as a function of the magnetic field measured in the presence of a direct current $I = 50$ $\mu$A. 
The gate voltage serves as a discrete parameter and is varied from $V_g = -0.2$ V (top, lowest density) to $+1.2$ V (bottom, highest density) in steps of 0.1 V. 
Traces are vertically offset by 0.2 (bottom to top). 
All traces show HIRO, with maxima (marked by 1, 2, 3 at the bottom trace) moving towards lower $B$ with increasing $V_g$ (density). 
Concurrently, the HIRO amplitude gradually increases at low $V_g$, but then suddenly jumps at $V_g \approx 0.3$ V (cf. thick line). 
Beyond this gate voltage, HIRO persist down to very low $B < 0.01$ T. 
As shown below, this abrupt enhancement of HIRO coincides with the onset of SB2 population which is accompanied by a significant increase of the quantum lifetime of SB1 electrons.

To obtain the electron density for the individual subbands, $n_1$ and $n_2$, we exploit the period of MISO visible in \rfig{fig1} as fast oscillations. 
MISO are described by \citep{raikh:1994,averkiev:2001,raichev:2008,dmitriev:2012}
\be
\delta \rho/\rho_0 = (2\ttr/\ttr_{12})\lambda_1\lambda_2\cos \left (2\pi\Delta/\hbar\oc\right )\,,
\label{eq.miso}
\ee
where $\Delta$ is the inter-subband separation, $\ttr_{12}$ is the inter-subband transport scattering time, and $\lambda_1$ ($\lambda_2$) is the Dingle factor of SB1 (SB2) electrons. 
The $N$-th MISO maximum thus occurs when
\be
\frac{\Delta}{\hbar\oc} = \frac {h}{e B}\frac {\delta n}{2}
= N\,,
\label{eq.N}
\ee
where $\delta n \equiv n_1 - n_2$.

In \rfig{fig2}(a) we show $N$ as a function of $1/B$ for different $V_g$ from 1.2 V (bottom) to 0.3 V (top), in steps of 0.1 V. 
As prescribed by \req{eq.N}, all data sets exhibit linear dependencies. 
From the slope of these dependencies we obtain $\delta n$ and the subband populations as
\be
n_{1,2} = \frac {n \pm \delta n}2\,,
\label{eq.n12}
\ee
where $+$ ($-$) corresponds to $n_1$ ($n_2$) and the total density $n$ is found from the Hall resistivity $\rho_H = B/ne$. 
The obtained values of $n_1$ (squares) and $n_2$ (circles) are presented in \rfig{fig2}(b) as a function of $n$. 
For $n > n^\star\approx 1.2 \times 10^{11}$ cm$^{-2}$, $n_1$ shows little change, remaining close to $n^\star$ \citep{note:comp}, i.e. the change of $n$ (dashed line) occurs due to increasing $n_2$ only. 
These findings conform with earlier studies of such heterostructures \citep{muraki:2000,nuebler:2012,dorozhkin:2016}. 
In particular, calculations \citep{nuebler:2012} revealed a strong spatial separation of the wave functions of SB1 and SB2 electrons. 
At high $V_g$, SB1 electrons are located near the front interface of the quantum well (closer to the donor layer), while SB2 electrons reside near the back interface (closer to the gate).
This fact explains why $n_2$ increases at the same rate as $n$ while $n_1$ remains essentially unchanged.

\begin{figure}[t]
\includegraphics{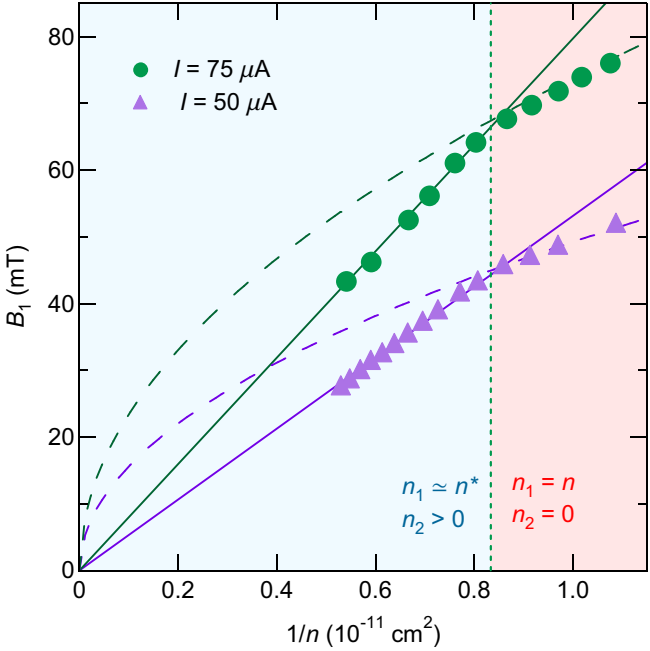}
\vspace{-0.1 in}
\caption{(Color online)
Magnetic field $B_1$ vs $1/n$ for $I=75$ $\mu$A (circles) and $I=50$ (triangles).
The vertical dotted line marks the onset of SB2 population.
Solid and dashed lines are calculated using \req{eq.b1m} with $n_1 = n^\star$ and \req{eq.b1}, respectively.
}
\vspace{-0.2 in}
\label{fig3}
\end{figure}

We next notice that even when SB2 is populated, HIRO remain periodic in $1/B$. 
This observation alone suggests that HIRO still originate predominantly from transport within a single subband. 
In \rfig{fig3} we present $B_1$, extracted from the data obtained at $I=75$ $\mu$A (circles) and $I=50$ (triangles), as a function of $1/n$. 
Both dependencies reveal two distinct regimes separated by a kink at $n \approx n^\star$. 
For the single-subband case, $n < n^\star$, one finds $B_1 \propto n^{-1/2}$, in agreement with \req{eq.b1} (dashed lines). 
At $ n > n^\star$, both dependencies change to $B_1 \propto 1/n$ (cf. solid lines), reflecting that HIRO stem from backscattering of SB1 electrons only. 
Indeed, while the Hall field is determined by the total density $n$, the cyclotron diameter of electrons in SB1 is controlled by $n_1$. 
As a result, the HIRO frequency in this regime needs to be modified as
\be
B_1 = \frac{\sqrt{8\pi n_1}}{n}\frac {\m}{e^2} j\,.
\label{eq.b1m}
\ee
Since $n_1 \approx {\rm const}$ at $n > n^\star$, one readily concludes that $B_1 \propto 1/n$. 
We can confirm the above picture by estimating $n_1$ from the HIRO frequency, employing \req{eq.b1} at $n < n^\star$ and \req{eq.b1m} at $n > n^\star$. 
The obtained values of $n_1$ [marked by crosses in \rfig{fig2}(b)] are in excellent agreement with $n$ at $n < n^\star$ and with $n_1$ obtained from \req{eq.n12} at $n > n^\star$.

\begin{figure}[t]
\includegraphics{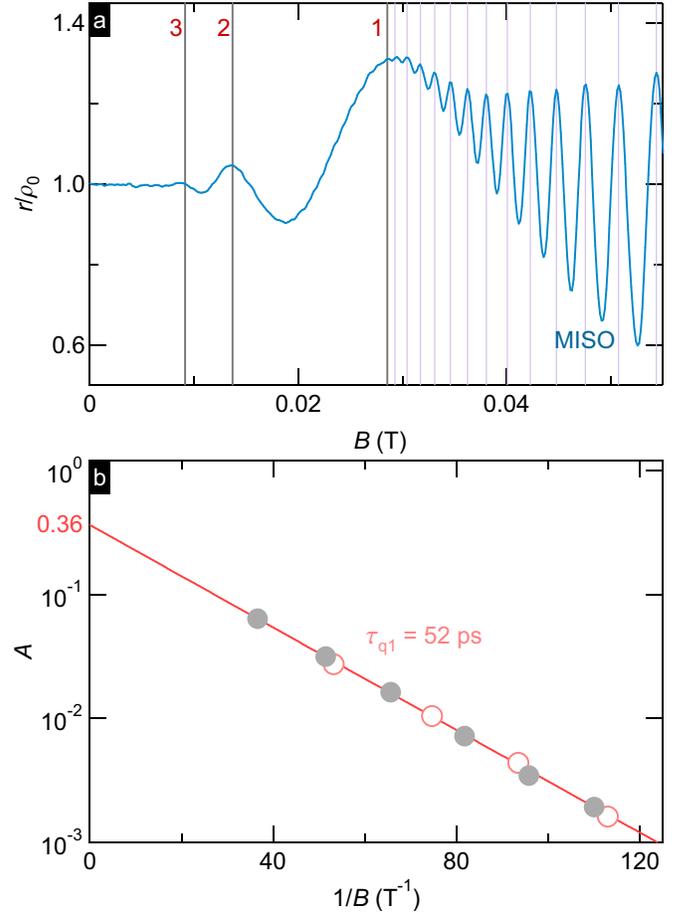}
\vspace{-0.1 in}
\caption{(Color online)
(a) Differential resistivity $r$, normalized to $\rho_0$, vs $B$ measured at $I = 50$ $\mu$A and at a gate voltage $V_g$ = 1.2 V ($n = 1.88\times10^{11}$ cm$^{-2}$). 
Thick lines mark HIRO maxima corresponding to $B/B_1 = 1, 2$, and 3 \citep{note:bessel}. 
Thin lines mark MISO maxima.
(b) Reduced HIRO amplitude $A = (\pi/16)(|\delta r_{\max}|/r_0)$ [see \req{eq.hiro}] as a function of $1/B$ on a log-linear scale for $n = 1.37\times10^{11}$ cm$^{-2}$ (filled circles) and $n = 1.88\times10^{11}$ cm$^{-2}$ (open circles). The fit (straight line) to $(\tau/\tpi)\exp(-2\pi/\oc\tqa)$ yields $\tqa \approx 52$ ps and $\tau/\tpi \approx 0.36$.
}
\vspace{-0.2 in}
\label{fig4}
\end{figure}
\begin{figure}[t]
\includegraphics{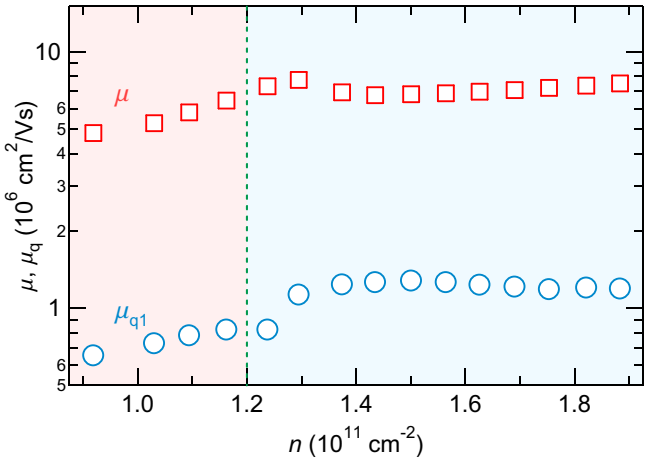}
\vspace{-0.1 in}
\caption{(Color online)
Mobility, $\mu$ (squares), and quantum mobility of SB1 electrons, $\mqa$ (circles) \citep{note:muq}, as a function of $n$.
}
\vspace{-0.2 in}
\label{fig5}
\end{figure}

Next, we take a closer look at HIRO in the two-subband regime. 
A representative trace is shown in \rfig{fig4}(a) displaying $r/\rho_0$ vs $B$ measured at $I = 50$ $\mu$A and $n = 1.88\times10^{11}$ cm$^{-2}$. 
The positions of the thick vertical lines, which pass through the HIRO maxima, are calculated using \req{eq.b1m}. 
As noticed earlier, HIRO persist down to a very low $B \approx 0.0075$~T, suggesting large quantum lifetime of SB1 electrons $\tqa$. 
To access its value, we construct a Dingle plot which is shown in \rfig{fig4}(b). 
The reduced HIRO amplitude $A = (\pi/16)(|\delta r_{\max}|/r_0)$ [see \req{eq.hiro}] is plotted as a function of $1/B$ on a log-linear scale for $n = 1.37\times10^{11}$ cm$^{-2}$ (filled circles) and $n = 1.88\times10^{11}$ cm$^{-2}$ (open circles). 
Remarkably, both data sets collapse on the same line indicating that, once SB2 is populated, further increase of $n_2$ has negligible effect on $A$. 
The amplitude $A$ exhibits the expected exponential dependence and the fit (solid line) yields $\tqa \approx 52$ ps. 
This value translates to a quantum mobility of $\mqa\equiv e\tqa/\m \approx 1.3\times 10^{6}$ cm$^2/Vs$, which, to our knowledge, is the largest value ever reported in any system \cite{note:46ps,shi:2016a}. 
The intercept of the fit with $1/B = 0$ gives $\tau/\tpi \approx 0.36$, implying that backscattering of SB1 electrons provides an essential fraction of the total transport scattering rate. 
The quantum mobility $\mqa(n)$ is shown in \rfig{fig5} together with the transport mobility $\mu(n)$ measured at $B=0$.

The overlapping Dingle plots in \rfig{fig4}(b) indicate that both $\tpi$ and $\tq$ do not vary in the two-subband regime. 
It is thus reasonable to expect that the transport mobility $\mu_1$ of SB1 electrons also stays constant at $n > n^\star$. 
This assumption enables us to estimate mobility of the second subband $\mu_2$. 
Indeed, using the relations $\mu n=\mu_1 n_1+\mu_2 n_2$ \citep{note:mu2}, $n_1\approx n^\star$, and $n_2\approx n-n^\star$, we find $\mu_2(n)\approx [n\mu(n)-n^\star\mu(n^\star)]/(n-n^\star)$, where $\mu_1\approx \mu(n^\star)\simeq 8 \times 10^6$ cm$^2$/Vs. 
This procedure yields $\mu_2 \approx 7 \times 10^6$ cm$^2$/Vs at $n_2 = 0.75 \times 10^{11}$ cm$^{-2}$, i.e., $\mu_2 \approx \mu_1$ even though $n_2 < n_1$.

With the knowledge of $\tqa$, we can also estimate the quantum lifetime of SB2 electrons $\tqb$, by employing the fact that the MISO amplitude decays as $\lambda_1\lambda_2 = \exp(-\pi/\oc\tqa)\exp(-\pi/\oc\tqb)$, see \req{eq.miso}.
The corresponding Dingle analysis yields $\tqb \approx 14$ ps ($\mqb \approx 0.35 \times 10^6$ cm$^2$/Vs) at $n_2 = 0.75 \times 10^{11} \text{cm}^{-2}$ \cite{note:miso12}. 
The low value of $\tqb$ readily explains why no HIRO from SB2 electrons \big[that would decay as $\exp(-2\pi/\oc\tqb)\ll\exp(-2\pi/\oc\tqa)$\big] is detected.
Indeed,  HIRO are barely visible at the lowest $n = n_1$ (c.f. top trace in \rfig{fig1}) even though $\mqa$ is twice as high.

We next discuss possible reasons why $\mqb \approx 0.3\mqa$ (at the highest $n$), which is rather surprising in view of $\mu_2 \approx \mu_1$. 
While at the highest $n$, $n_1$ is 50\% higher than $n_2$, this fact alone cannot account for the observed difference between $\mqb$ and $\mqa$.
In modulation-doped structures, large $\mu_i/\mqi$ ratio reflects dominance of small-angle scattering originating from donors separated by a spacer of thickness $d_1$ ($d_1 = 66$ nm in our sample). 
Based on our estimates above, we obtain, at the highest $n$, $\mu_1/\mqa \approx 6$ and $\mu_2/\mqb \approx 20$, which indicates that SB2 electrons experience a much stronger small-angle scattering.
However, one can expect just the opposite relation since (i) SB2 electrons are further away from the donor layer and (ii) SB1 electrons should effectively screen the random potential from this layer. 
For the same reason, a strong (about 50\%) enhancement of $\tqa$ in the two-subband regime \citep{note:lu,lu:1998}, shown by circles in \rfig{fig5}, is also puzzling, as it can not be attributed to an additional screening of the donor random potential by SB2 electrons. 

To explain our findings, one apparently needs to consider another kind of distant scatterers, located closer to the back side of the quantum well. 
One possible source of such scatterers is the doped GaAs quantum well (serving as a back gate), located at a distance $d_2 = 800$ nm.
Since the gate is much farther away than the donor layer ($d_1/d_2 \sim 0.1$), it has a negligible impact on mobility.
However, it can still limit the quantum mobility which is dominated by small-angle scattering \citep{qian:2017b}; with a random charge density of $n_2  = 1\times 10^{11}$ cm$^{-2}$, it would limit the quantum mobility of SB2 electrons to $\mqb \approx (4ed_2/\hbar)/\sqrt{2\pi n_2} \approx 0.6 \times 10^6$ cm$^2$/Vs \citep{dmitriev:2012}.
This estimate is not unreasonable \citep{note:dl}, considering that other scattering sources further diminish $\mqb$.
Almost complete screening of the gate random potential by SB2 electrons should then lead to an enhancement of $\mqa$ observed at the onset of the population of the second subband.
Similarly, SB2 electrons should partially screen the long-range component of the potential created by background impurities leading to further enhancement of $\mqa$ in the two-subband regime.
The above discussion demonstrates that combined analysis of the various magnetooscillations in multisubband heterostructures provides rich information on the characteristics of the disorder potential, which may help to improve our understanding and allow to technologically control the role of particular scattering sources.

In summary, we have studied nonlinear magnetotransport in a 60 nm-wide GaAs/AlGaAs quantum well equipped with an \emph{in situ} grown back gate for tuning the carrier density $n$. 
An analysis of HIRO and MISO frequencies indicates that the occupation of the second subband is triggered at $n = n^\star \approx 1.2 \times 10^{11}$ cm$^{-2}$.
At ($n < n^\star$) $n > n^\star$, we find that the HIRO frequency increases as $n^{-1/2}$ ($n^{-1}$), indicating that the observed HIRO originate from backscattering of SB1 electrons at all densities studied. 
Remarkably, when SB2 becomes populated, HIRO suddenly become much more pronounced. 
Our analysis shows that this HIRO enhancement is due to an abrupt increase of quantum lifetime of SB1 electrons. It approaches a record value of 52 ps in the two-subband regime of transport. 
The enhancement likely originates from additional screening of the disorder potential by SB2 electrons.

\begin{acknowledgments}
We thank J.~Falson for assistance with the experimental setup and M.~Borisov for performing the Dingle analysis of MISO. 
The work at the University of Minnesota was funded by the U.S. Department of Energy, Office of Science, Basic Energy Sciences, under Award \# ER 46640-SC0002567.
Q.S. acknowledges The University of Minnesota Doctoral Dissertation Fellowship.
I.D. acknowledges financial support from the German Research Foundation (DFG grant No. DM1/4-1). 
V.U. and J.H.S acknowledge financial support from the GIF.
\end{acknowledgments}

\vspace{-0.1 in}

\begin{thebibliography}{56}
\bibitem[{not({\natexlab{a}})}]{note:times}
\bibinfo{note}{The rate of scattering on angle $\theta$ can be expressed in
  terms of angular harmonics as $\tau_\theta^{-1} = \sum\tau_n^{-1}e^{i
  n\theta}$. In this notation, $\tq^{-1}=\tau_0^{-1}$,
  $\ttr^{-1}=\tau_0^{-1}-\tau_1^{-1}$, and $\tpi^{-1}= \sum(-1)^n\tau_n^{-1}$.}
\bibitem[{not({\natexlab{g}})}]{note:muq}
\bibinfo{note}{$\mqa$ was obtained from the fit of the $r(B)$ data with the
  theoretical expression shown in \citep{note:bessel}.}
\bibitem[{not({\natexlab{h}})}]{note:mu2}
\bibinfo{note}{While $\mu_{1,2}$ include both intra- and inter-subband
  scattering, the latter is relatively weak \citep{note:miso12}.}
\bibitem[{not({\natexlab{k}})}]{note:dl}
\bibinfo{note}{Application of the same procedure to the donor layer would
  result in underestimated $\mqa$, likely because it ignores screening of
  ionized donors by $X$-electrons in AlAs layers surrounding GaAs doping well.}
\end{thebibliography}

\end{document}